\documentclass[twocolumn]{aastex62}
\usepackage{amsmath}
\usepackage{amssymb}
\usepackage{graphics}
\usepackage{hyperref}
\usepackage{soul}

\newcommand\plottwocropped[2]{%
	\centering 
	\leavevmode 
	\includegraphics[trim={3.5cm 1.7cm 2cm 1cm},clip,scale=0.25]{#1}
	\hfil
	\includegraphics[trim={3.5cm 1.7cm 2cm 1cm},clip,scale=0.25]{#2}
}
\graphicspath{{./}{}}

\received{January 1, 2018}
\revised{January 7, 2018}
\accepted{\today}
\submitjournal{ApJ}

\shorttitle{Estimation of CME plasma properties}
\shortauthors{Mondal et al.}

\begin{document}

\title{Estimation of the physical parameters of a CME at high coronal heights using low frequency radio observations }

\correspondingauthor{Surajit Mondal}
\email{surajit@ncra.tifr.res.in}

\author{Surajit Mondal}
\affil{National Centre for Radio Astrophysics, \\
Tata Institute of Fundamental Research, \\
Pune-411007, India}

\author{Divya Oberoi}
\affil{National Centre for Radio Astrophysics, \\
Tata Institute of Fundamental Research, \\
Pune-411007, India}


\author{Angelos Vourlidas}
\affil{John Hopkins University Applied Physics Laboratory \\ Laurel, United States of America}



\begin{abstract}
Measuring the physical parameters of Coronal Mass Ejections (CMEs), particularly their entrained magnetic field, is crucial for understanding their physics and for assessing their geo-effectiveness. At the moment, only remote sensing techniques can probe these quantities in the corona, the region where CMEs form and acquire their defining characteristics. Radio observations offer the most direct means for estimating the magnetic field when gyrosynchontron emission is detected. 
In this work we measure various CME plasma parameters, including its magnetic field, by modelling the gyrosynchrotron emission from a CME. 
The dense spectral coverage over a wide frequency range provided by the Murchison Widefield Array (MWA) affords a much better spectral sampling than possible before.
The MWA images also provide much higher imaging dynamic range, enabling us to image these weak emissions.
Hence we are able to detect radio emission from a CME at larger distances ($\sim 4.73 R_\odot$) than have been reported before.
The flux densities reported here are amongst the lowest measured in similar works.
Our ability to make extensive measurements on a slow and otherwise unremarkable CME suggest that with the availability of data from the new generation instruments like the MWA, it should now be possible to make routine direct detections of radio counterparts of CMEs.

\end{abstract}



\section{Introduction} \label{sec:intro}

Coronal Mass Ejections (CMEs) are eruptions of magnetized plasma from the solar atmosphere, representing the most energetic explosions in the solar system.
Although the details of their initiation and early evolution are not well understood yet, there is a general consensus that these explosions are driven by the magnetic fields \citep{aschwanden04book}.
Magnetic field measurements of CMEs (both inside the CME and at the shock front) can hence serve as powerful constraints for CME initiation and evolution models.
The geo-effectiveness of a CME is also determined by its magnetic field \citep[e.g.][]{plunkett2000}.
While their importance is well recognized, remote measurements of CME magnetic fields are challenging and only a handful of examples exist in the literature.

Several techniques have been used in the past to estimate the magnetic field at the shock front of a CME. 
Some of the more popular techniques that have been used to estimate the average magnetic field at the shock front are using band splitting of type II bursts observed in the  solar radio dynamic spectrum \citep[e.g.][and many others]{smerd75,gary1984,cunha-silva15,kumari17a,mahrous18}; circular polarisation of moving type IV bursts \citep{raja14,kumari17b}; and the standoff distance technique using extreme ultraviolet and optical images\citep{gopalswamy11,gopalswamy12,poomvises12}. 
\citet{susino15} developed and successfully applied a technique for estimating the spatially varying magnetic field at the shock front under plausible assumptions.
However, none of these techniques can be used to determine the magnetic field entrained in the CME. Another common limitation of these techniques is that they do not provide any information about the non-thermal particle distribution in the CME.
It is well known that shocks associated with CMEs are very efficient particle accelerators \citep{ackermann17}.
Hence it is important to get quantitative estimates of the energy spectrum of the accelerated particles and its time evolution.

\citet{bastian01} demonstrated that it is possible to estimate both the local magnetic field and non-thermal particle distribution inside a CME using multi-frequency radio observations by modelling the spectrum of gyrosynchrotron emission from these energetic particles.
Observations in the metric and decimetric wavelength are best suited for detecting coronal gyrosynchrotron emission \citep{bastian1997}.
Since this emission is completely determined by local plasma properties, radio maps can provide spatially resolved information about the parameters of the CME and the coronal plasma.
Due to the rich information content of the CME gyrosynchrotron spectrum, significant efforts have been made towards such studies.
Despite this, only a few successful detections of gyrosynchrotron emission from CMEs have been reported in the literature  \citep[e.g.][]{bastian01,maia07,Tun13,bain14,carley17}.

Here we present a detailed study of radio emission from CME plasma, based on data from the Murchison Widefield Array \citep[MWA;][]{Lonsdale09,tingay13}. 
We are able to fit gyrosynchrotron model to the observed spectra at multiple locations and times, and also detect radio emission from the CME at the largest heliocentric distance to date.
We estimate both the CME magnetic field and the nonthermal electron distribution from the spectral fits.
We also find significant evidence for variability in the observed spectra.
Section \ref{Sec:Obs} details the observations and the data analysis procedure, while the results from radio imaging and spectral modeling are presented in Sec. \ref{Sec:Results}. 
Section \ref{Sec:Dicsussion} presents a discussion on the morphology and emission mechanisms involved, including a comparison with earlier reports in the literature; and Sec. \ref{Sec:Conclusion} concludes the paper.

\section{Observation and data analysis}
\label{Sec:Obs}

The observations presented here were made on November 4, 2015.
This day is charaterized by a ``high" level of solar activity\footnote{https://www.solarmonitor.org/?date=20151104}.
Eight active regions were present on the visible disc of the Sun.
The Space Weather Predicton Center (SWPC) event list reports numerous radio and X-ray events including three GOES M-class flares.
The Coordinated Data Analysis Workshops (CDAW) CME catalogue lists 11 CMEs for this day\footnote{CDAW catalog, \url{https://cdaw.gsfc.nasa.gov/CME_list/UNIVERSAL/2015_11/univ2015_11.html}}.

\begin{figure}
    \centering
    \includegraphics[trim={1.5cm 1cm 1.5cm 1cm},clip,scale=0.62]{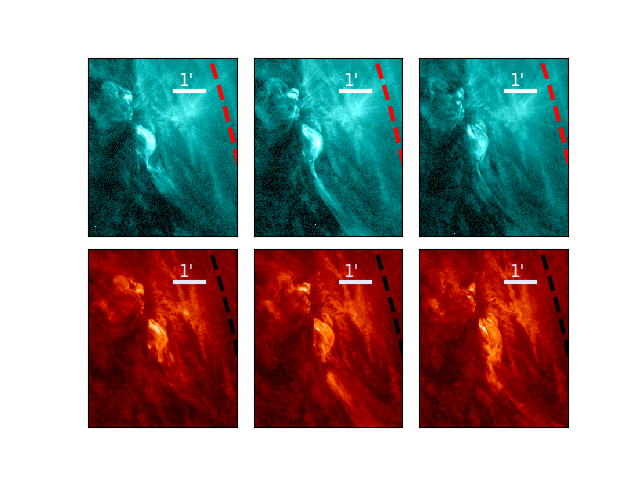}
    \caption{The top and bottom panels show some example images during the eruption phase from AIA 131 \AA\ and 304 \AA\ respectively.
    Left, middle and right panels correspond to the available frame nearest to 01:35:54 UT, 01:40:54 UT and 01:45:54 UT respectively. The spatial scale is shown in each panel by a thick line. The dotted line shows the solar limb.}
    \label{fig:AIA_images}
    
\end{figure}

\begin{figure*}
    \centering
    \includegraphics[trim={1.2cm 0 0 0},clip,scale=0.7]{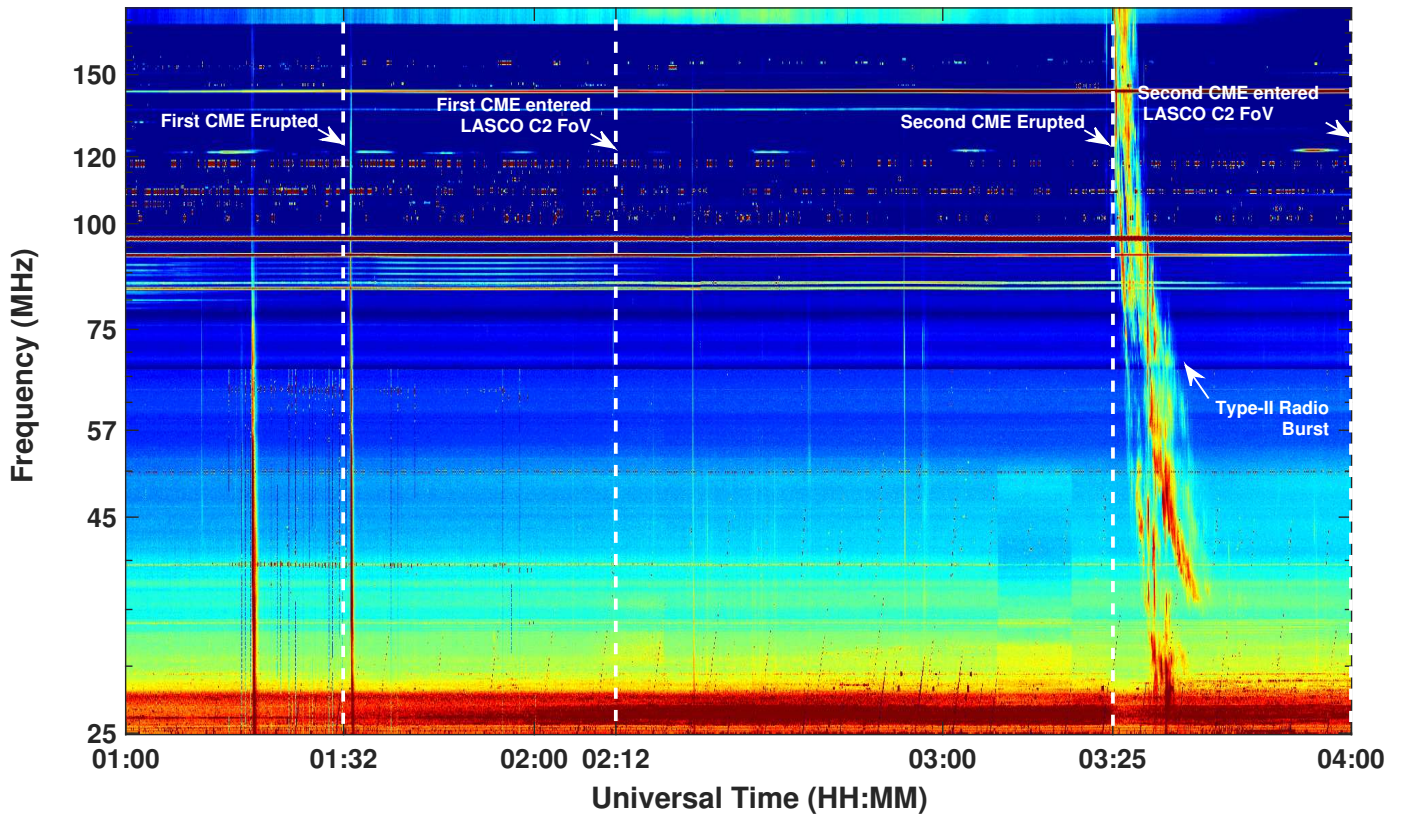}
    \caption{Dynamic spectrum from the Learmonth Solar Radio Spectrograph spanning the observation interval of interest. The type-II burst associated with the later CME is clearly seen. 
    The horizontal lines arise due to persistent radio frequency interference. A timeline of different events relevant to this work is marked by white dashed lines. 
    }
    \label{fig:dynamic_spectrum}
\end{figure*}

The CME of this study first appeared in the field of view (FOV) of the Large Angle and Spectrometric Coronagraph \citep[LASCO;][]{brueckner1995} C2, onboard the Solar and Heliospheric Observatory \citep[SOHO;][]{domingo1995} at 02:12 UT.
The CDAW catalog radial speed of this CME is 442 $km \; s^{-1}$.
Assuming a constant speed since initiation we estimate an upper limit for the time of eruptions to be about 01:38 UT. 
Based of this, we associate this CME with the eruption which took place at NOAA 12445 (N16W82) around 01:32 UT.
This eruption was evident both in the hot (e.g. 131 \AA\, sensitive to plasma temperature of ~12 MK) and cool channels (e.g. 304 \AA\, sensitive to plasma temperature of 0.05 MK) of the Atmospheric Imaging Assembly \citep[AIA;][]{lemen2012} onboard the Solar Dynamics Observatory \citep[SDO;][]{pesnell2012}. 
Figure \ref{fig:AIA_images} shows some example AIA images at 131 \AA\ and 304 \AA\ while this eruption is in progress.
This CME was observed in the LASCO C2 FOV from 02:12 UT to about 05:00 UT. We refer to this CME as the ``first CME".
Another eruption took place from the same active region at 03:25 UT.
It was accompanied by a M1.9 X-ray flare and a type-II radio burst, and first appeared in the LASCO C2 FOV at 04:00 UT.
This event has been studied in detail by \citet{kumari17b} and \citet{ying18} and is referred to as the ``second CME" in  Fig. \ref{fig:dynamic_spectrum}, which shows the timeline of the relevant events between 01:30-04:00 UT using a radio dynamic spectrum.
{The second CME is not discussed any further here. Some other CMEs also erupted on this day.
These events have been discussed in detail in \citet{cairns2019} and will not be discussed here.}


The emphasis of this work is on analysis of metrewave radio data.
These data come from the MWA and cover the time range from 03:03 UT to 03:35 UT.
As discussed in Section \ref{Subsec:radio-imaging}, although this timerange includes the second CME, the radio emission studied here is not related to it.
The observations were carried out in 12 frequency bands, each of 2.56 MHz bandwidth, and centred close to 80, 89, 98, 108, 120, 132, 145, 161, 179, 196, 217, and 240 MHz.
The time and spectral resolution of the data were 0.5 s and 40 kHz, respectively.
For context, the dynamic spectrum obtained from the Learmonth Solar Radio Spectrograph spanning this period is shown in Fig. \ref{fig:dynamic_spectrum}.
Imaging was done using the Automated Imaging Routine for Compact Arrays for the Radio Sun \citep[AIRCARS;][]{mondal19}.
The final images had a spectral resolution of 2 MHz.
The time resolution ranges from 0.5-10 s.
The objective of this study was to detect the gyrosynchrotron emission from CME plasma, which is known to be significantly fainter than the quiescent solar emission \citep[e.g.][]{bastian01}, in presence of a noise storm which is at least an order of magnitude brighter than the quiescent Sun.
AIRCARS performance was therefore tuned to provide high dynamic range images.
Typical dynamic range of images used in this study is $\sim$13,000.
These images were flux calibrated following the methods described in \citet{oberoi17} and \citet{mohan17}. 
Flux calibration is not yet available for frequencies below 100 MHz, hence they are not used for quantitative analysis.

\section{Results}
\label{Sec:Results}

\subsection{Radio imaging}
\label{Subsec:radio-imaging}

\begin{figure*}
    \centering
    \includegraphics[trim={7cm 2cm 5cm 7cm},clip,scale=0.39]{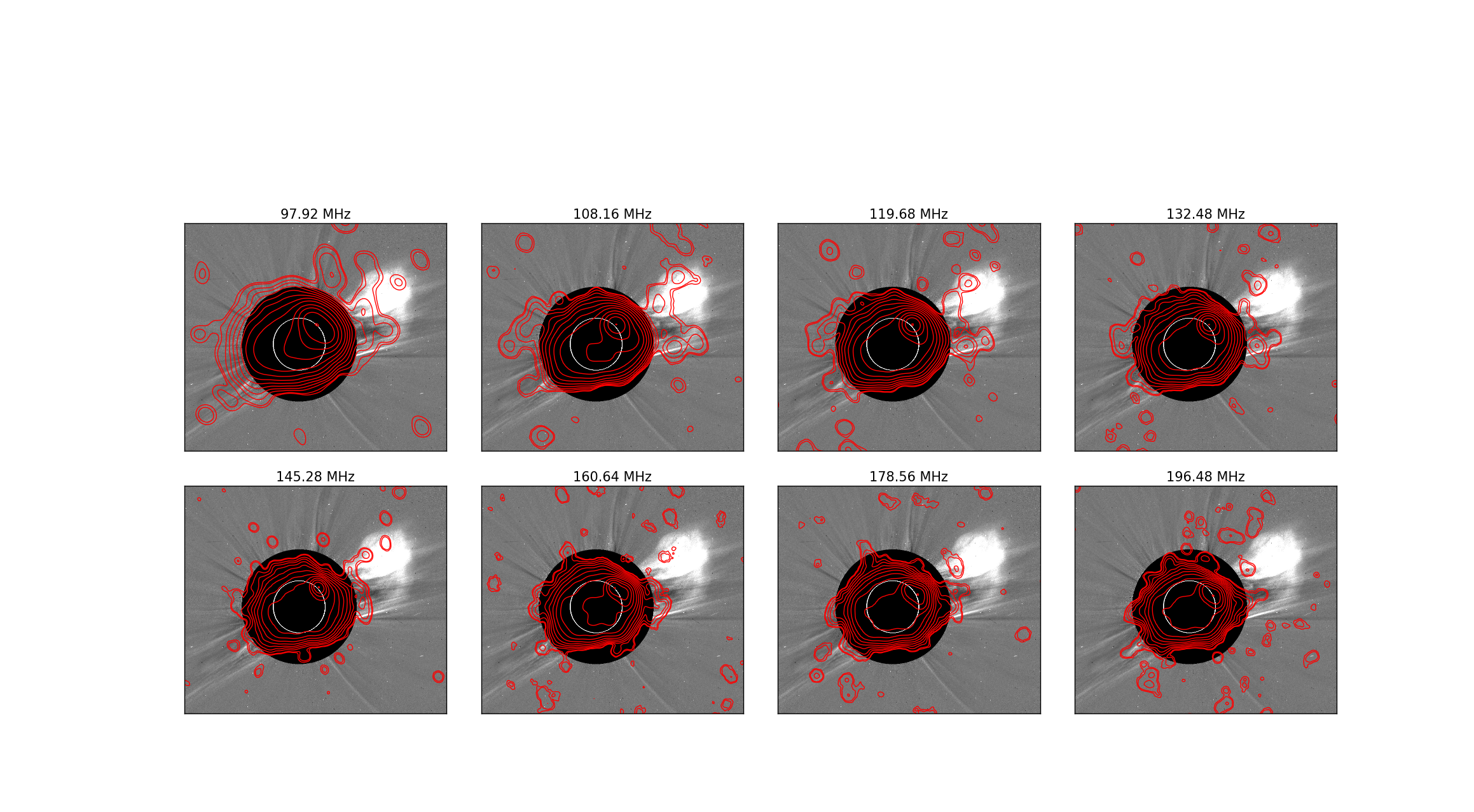}
    \caption{Radio contours at different frequencies at 03:32 UT overlaid on LASCO C2 image at 03:36 UT. 
    The central frequency for each image is mentioned.
    The contour levels start at 0.02\% of the peak, and increase in multiples of two.
    }
    \label{fig:same_time}
\end{figure*}

\begin{figure}
    \centering
    \plottwocropped{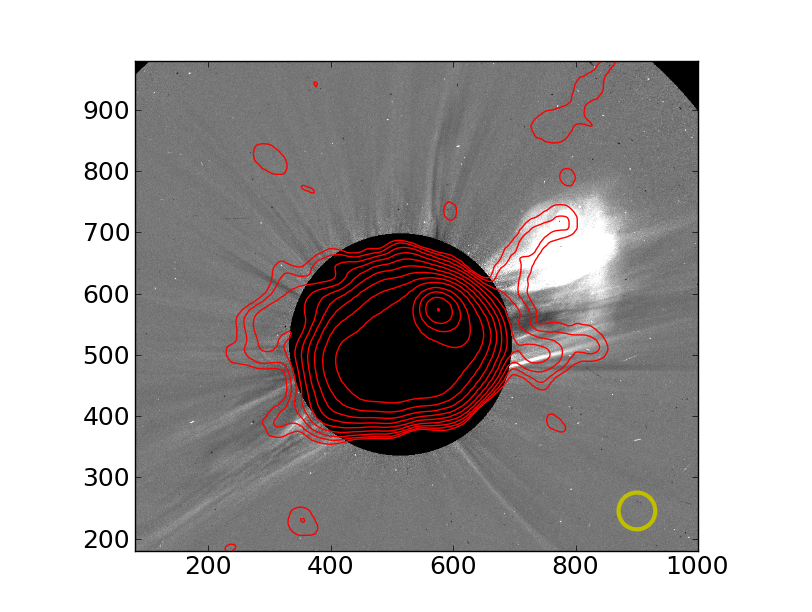}{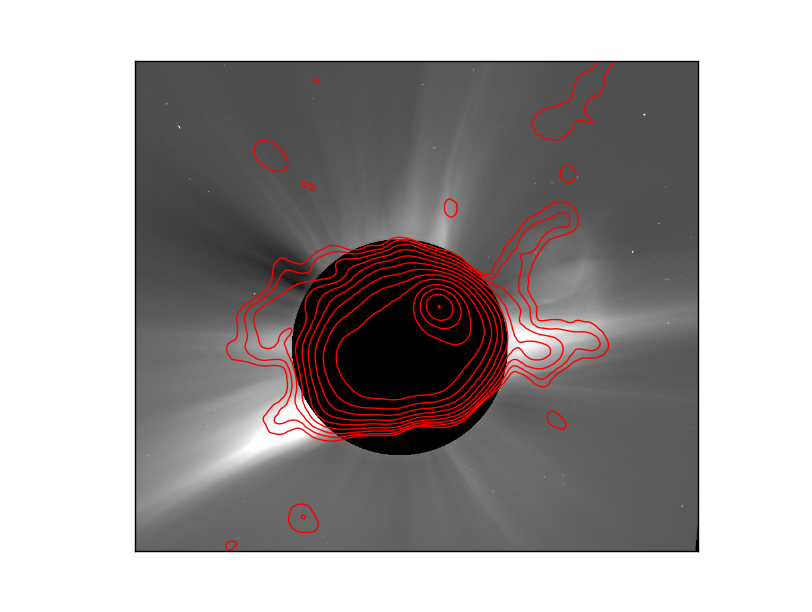}
    \caption{Contours of average normalised image (108-145 MHz) overlaid on the LASCO C2 difference image (left panel) and LASCO C2 image (right panel). The contour levels are at 0.0002, 0.0004, 0.0008, 0.0016, 0.0032, 0.0064, 0.0128, 0.0256, 0.0512, 0.124, 0.248, 0.496, 0.992 times the peak. The yellow circle in the lower right corner of the left panel shows the point spread function.}
    \label{fig:avg_normalised}
\end{figure}

Figure \ref{fig:same_time} shows the radio contours at 03:32 UT corresponding to different frequencies overlaid on LASCO C2 difference image at 03:36 UT.  
The lowest contour in each image is at 0.02\% of the peak and subsequent contour levels increase in multiples of 2.
The lowest contours have been chosen such that noise peaks are visible in all of the images.
It is also evident that many of these peaks are not correlated across different frequencies, but there is a region of extended emission which is seen in all of the images from 96 MHz to 146 MHz.
Not only is the morphology of this emission strongly correlated across neighboring frequencies, it is also seen to evolve systematically with frequency.
This leads us to believe that this faint extended emission feature has indeed been reliably detected in the images spanning the range from 98 MHz to 146 MHz.
The faint emission clearly extends to heliocentric distances $\gtrsim 2.3 R_\odot$ at all frequencies shown in Fig \ref{fig:same_time}. 
To confirm the presence of this faint radio emission, we smoothed the images from 108-145 MHz to a common resolution and then averaged them after normalising with their respective peak values. The contours of this average and normalised radio image are overlaid on the LASCO C2 difference image and LASCO C2 image in left and right panel of Fig. \ref{fig:avg_normalised} respectively. The last contour has been chosen so that it also shows the noise in the image. In the sky plane, the east protrusions are located near the white light streamers. In Section \ref{sec:radio_CME_discussion}, we give our reasons why we believe that the radio structure detected here at the west limb is related to the CME itself. Hence we suggest that we are also observing a `radio CME' even though the radio structures detected here do not show a clear circular bubble similar to the ones seen by \citet{bastian01} and \citet{maia07}.
We adopt this definition here, with the caveats expressed above.

The second CME entered the LASCO C2 FOV only at 04:00 UT and  based on radial speeds from the CDAW catalogue was at $\sim 1.2 R_\odot$ at 03:32 UT. Hence, we associate all emission at the west limb which arises from a heliocentric distance beyond 1.8 $R_\odot$ with the first CME.


It should be noted that although Fig. \ref{fig:dynamic_spectrum} shows the presence of the type II burst between 03:23-03:35 UT, it was present for a much smaller time interval in the frequencies of interest ($>98$ MHz) of this paper. During the times, when the type II burst was present within our observation band, the images were severely dynamic range limited, and we did not detect any feature except the coherent emission coming from the type II source.

\begin{figure}
    \centering
    \includegraphics[trim={15cm 10cm 1cm 2.5cm},clip,scale=0.42]{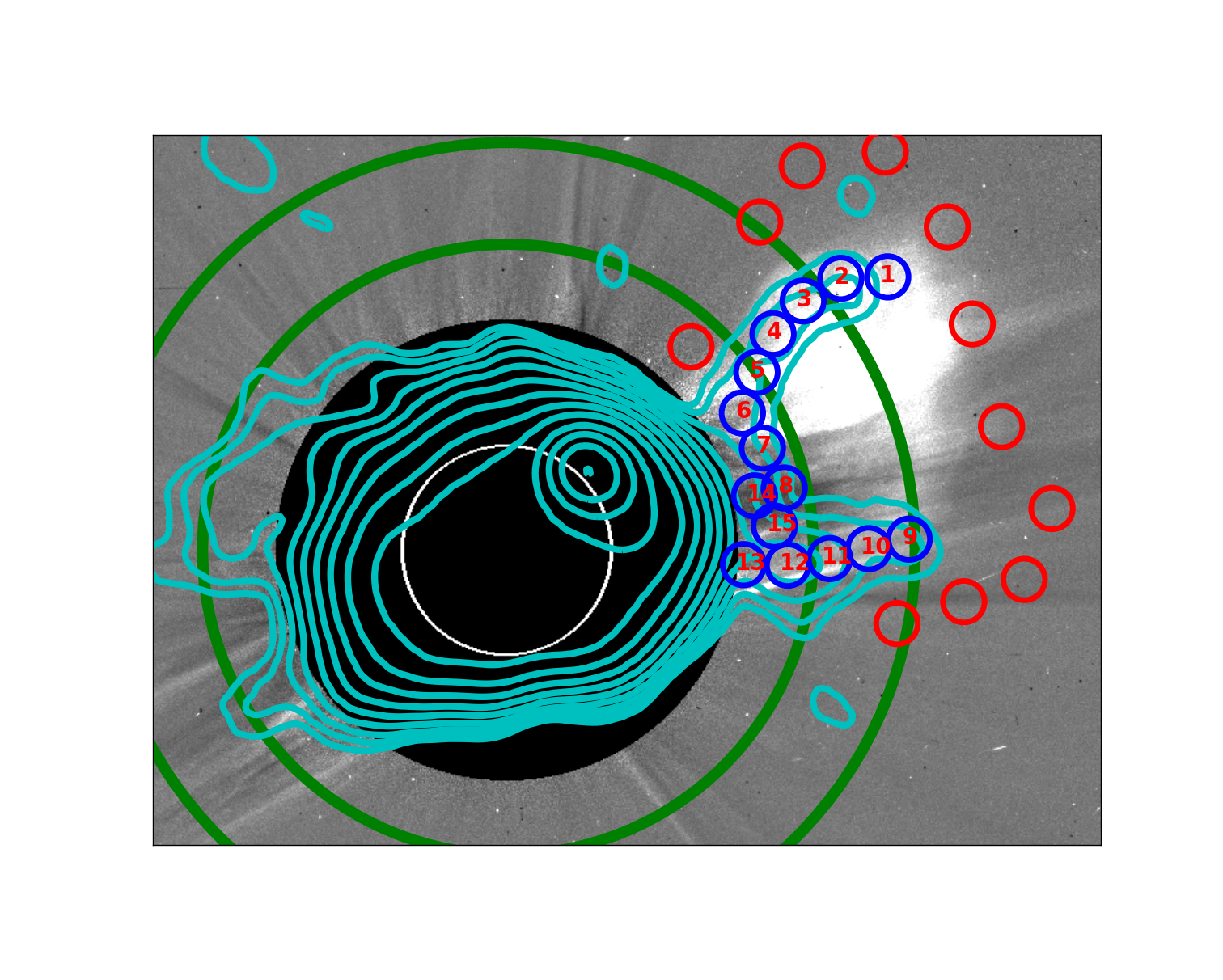}
    \caption{Regions where spectra have been extracted are marked in blue circles. The red circles indicate regions from where no solar or CME emission is expected, and their flux density is used for estimating measurement noise. 
    The green circles are drawn at a radii of 3 and 4 $R_\odot$ respectively. }
    \label{fig:region_spectra}
\end{figure}

For a quantitative analysis, the radio spectra for multiple regions were constructed at different times. The regions from which the spectra have been extracted are the same at all times.
These regions are marked by numbered blue circles in Fig. \ref{fig:region_spectra}.
The area of each of these regions is equal to the area of the point spread function (PSF) at 108 MHz.
The un-numbered red circles mark some of the regions where we do not expect any radio emission from the Sun or the CME plasma. 
The observed flux densities in these regions was used to estimate the uncertainty in each of the flux density measurements.
For quantitative analysis only those points have been used for which the flux density is greater than both $\mu+3 \sigma$ and $3 \alpha$, where $\mu$ and $\sigma$ are the median, standard deviation of flux densities measured in the red circles respectively, and $\alpha$ is the rms measured in a part of the map far away from the Sun.
$\mu$, $\sigma$ and $\alpha$ were calculated independently for each time and frequency slice. 
The errorbar shown in each point corresponds to the quadrature sum of max($\mu$, $\alpha$) and a systematic uncertainty in flux density estimates of 10\%, discussed in detail in \citet{oberoi17}.

\subsection{Modeling gyrosynchrortron spectra}
\label{Subsec:gryo-modeling}
\citet{bastian01, Tun13, bain14} have shown that radio emission from the core of the CME can arise from a gyrosynchrotron mechanism. 
We investigate if the same is true in this instance.
Gyrosynchrotron models involve many independent parameters \citep{ramaty1969}.
The gyrosynchrotron spectrum depends very sensitively on the local magnetic field vector, the number density of thermal electrons, and the number density and energy distribution of the relativistic electrons. 
The total volume of emission is also important as that determines the total number of electrons and also determines the level of self-absorption. 
Although the spatial distribution of these quantities is important for modelling the emission, for simplicity it is assumed that the emitting volume is homogeneous and the energetic electrons have an isotropic distribution.
In spite of this and assuming the simplest physically motivated distribution of the energetic electrons (a power law distribution between some $E_{min}$ and $E_{max}$ with a powerlaw index of $\delta$) the number of unknowns is 9; namely - area of emission, length along the line of sight (LOS), $E_{min}$, $E_{max}$, $\delta$, density of both thermal and non-thermal electrons, magnetic field strength and the angle between the magnetic field and the LOS.
This is further complicated by the fact that different parts of the spectra are sensitive to different parameters. 
For instance, the turnover point and the high frequency part of the spectrum are most sensitive to the magnetic field strength. There are also some degeneracies. For example, the same spectral peak height can be obtained by the combination of a larger magnetic field strength and smaller LOS angle or the other way around.
To break these degeneracies, it is important to obtain information from independent sources. 
In this particular instance, a Stokes V (circular polarization) spectrum can  break the degeneracy between magnetic field strength and LOS angle.
However, sufficient information to reliably constrain these parameters is not always available. 
Work is in progress to develop an appropriate polarization calibration algorithm for solar MWA images.
A heuristic approach to mitigate instrumental polarization leakage \citep{mccauley2019} has been remarkably successful, but is not applicable to the weak emissions being studied here.
Hence, the only recourse is to make plausible assumptions or informed guesses about the values of some the parameters.


In order to be in a regime where we can meaningfully constrain key model parameters of interest, we have restricted ourselves to using spectra with at least 6 flux density measurements.
The regions which satisfy this criterion are located at a heliocentric distance between $2.2-2.7R_\odot$.
Additionally, as described in the following text, we constrain some of these model parameters using independent measurements, make some reasonable simplifying assumptions, and set some of the model parameters to physically motivated constant values.
The electron density estimated from LASCO C2 polarised brightness map taken at 02:58 UT on the day of our observations is found to be $2\times 10^6\ cm^{-3}$ at $\sim2.5 R_\odot$, when measured at position angle of $\sim 295^\circ$ ccw.
The density of thermal electrons for all regions studied is fixed at this value as the range of heliocentric distances of these regions is approximately equal to the size of the regions themselves (PSF).
The local magnetic field is assumed to be perpendicular to the line of sight.
The non-thermal electron population is assumed to be isotropic, homogeneous and follow an energy distribution  described by a power law, $n_{nth}(E) \propto E^{-\delta}$ between some $E_{min}$ and $E_{max}$, where $E_{min}$ and $E_{max}$ are the lower and higher energy cutoffs of the power law, and $\delta$ the slope of the power law.
The number density of non-thermal electrons ($n_{nth}$) is set to $3\times 10^4$ $cm^{-3}$ (about 1.5\% of the thermal electron density), and $E_{max}$ to 10 MeV. 
As it turns out, the exact value of $E_{max}$ is not important as the large values of $\delta$, observed in these spectra, imply that there are few electron populating the spectrum close to $E_{max}$. 
$E_{min}$ was varied between 1--10 keV by hand till a satisfactory fit was obtained, and was held fixed at this value during the actual minimization procedure.
This choice is also motivated by the fact that when an enhancement is observed in the higher energy bands of RHESSI ($\sim 100$keV) , the low energy bands ($\sim 10$keV) always show an accompanying enhancement \citep{cheng2012}.

Spatially resolved spectroscopic X-ray imaging at the location of the nonthermal electrons can actually be used to constrain the non-thermal electron distribution. 
The emission from the electron population responsible for gyrosynchrotron emission in the CME core is, however, expected to be very faint, and is very difficult to detect with the current generation of X-ray instrumentation.
To the best of our knowledge, this has been successfully carried out only in one instance, when the CME erupted on the far side of the sun \citep{carley17}.
The bright X-ray emission from the loop foot point was hence occulted by the solar disc.
The consequent reduced imaging dynamic range requirement enabled \citet{carley17} to image the faint X-ray emission.
They found $E_{min}$ to be 9 keV. 
For this study, there is another reason as well why RHESSI data is insufficient. The regions being modeled lie at heliocentric distances of $\sim 2.5 R_\odot$, while RHESSI is sensitive only out to $\approx 1.8 R_\odot$\footnote{\url{https://hesperia.gsfc.nasa.gov/rhessi3/mission/mission-facts/index.html}}.

\begin{figure*}
    \centering
    \includegraphics[scale=0.4]{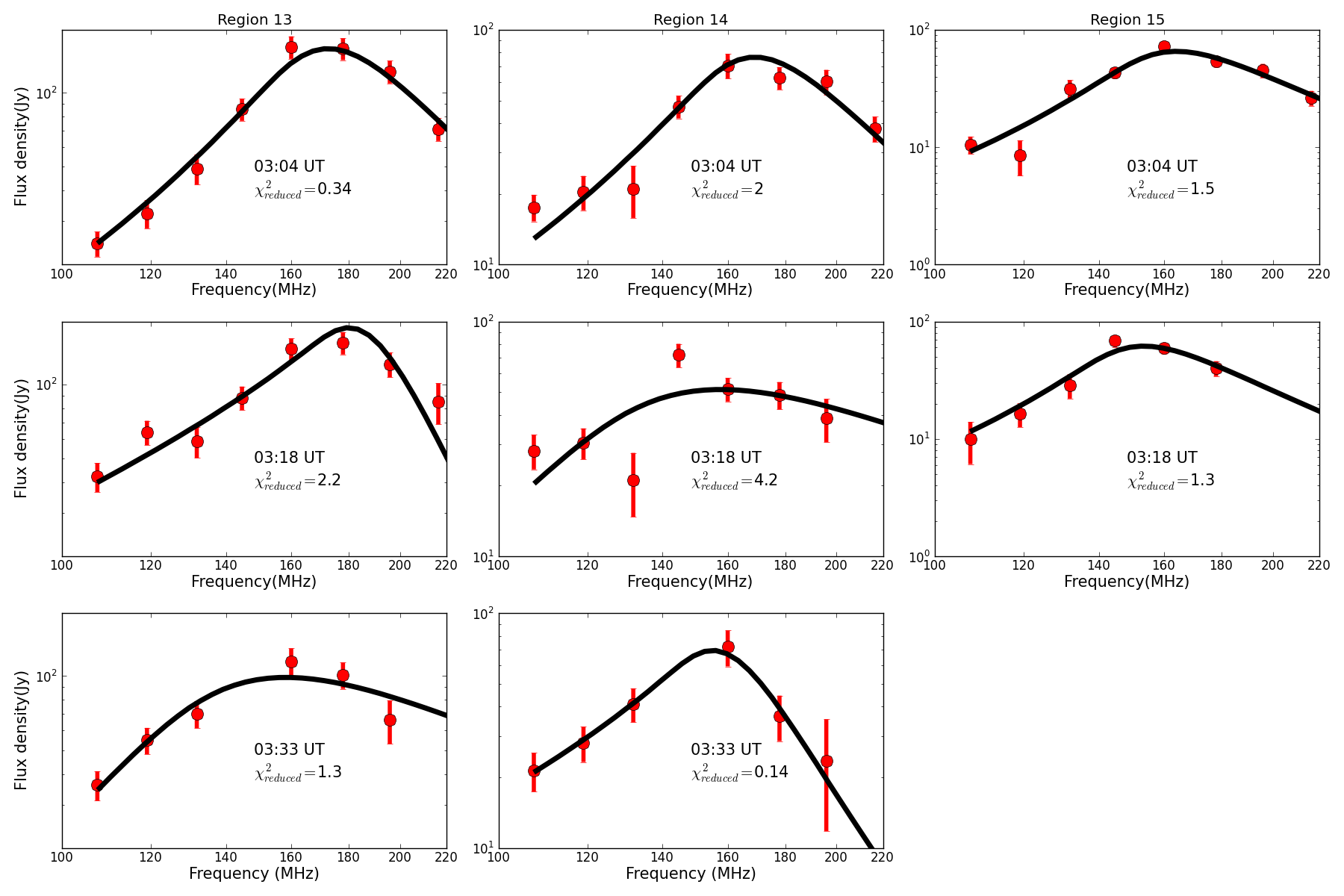}
    \caption{Gyrosynchrotron model fitted spectra. Row-wise, the first, second and third panel are for times 03:04, 03:18 and 03:33 UT. The region number of each spectrum is mentioned in each plot. The red points denotes the actual data while the black line denotes the fitted spectra.}
    \label{fig:model_spectra1}
\end{figure*}

\begin{figure*}
    \centering
    \includegraphics[scale=0.45]{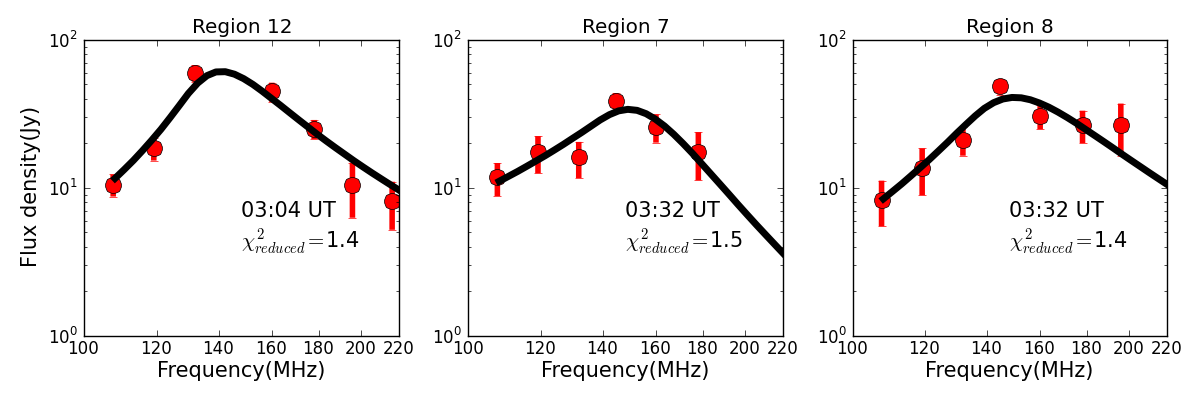}
    \caption{Gyrosynchrotron model fitted spectra. The region number and the time of the spectra and mentioned in each plot.}
    \label{fig:model_spectra2}
\end{figure*}

The depth along the LOS (L) was kept fixed at $3\times 10^{10}$ cm for fitting spectra for all regions except for Regions 7 and 8, where it needed to be changed to $4\times 10^{10}$ cm to obtain a satisfactory fit. 
These values of L were chosen because these are both close to the PSF size, and also the values used by previous authors for modeling gyrosynchrotron spectra.
The remaining three parameters, $\delta$, magnetic field (B) and the emission area (A) were the free parameters during the spectral fitting procedure.
For Region 14 at 03:18 UT, the emission area could not be constrained, it was hence fixed it to $6\times 10^{16}$ cm$^2$.
The gyrosynchrotron spectral modeling code by \citet{fleishman10} was used for modeling the spectra.
The results from spectral modeling are listed in Table \ref{tab:model_spectra} and the corresponding spectra are shown in Figs. \ref{fig:model_spectra1} and \ref{fig:model_spectra2}.
Using the fitted spectra, we also find that the observed powerlaw index in the rising parts of the spectra lie between 2.8-7, with the median powerlaw index being 3.8 which is much steeper than 2 that is expected from optically thick free-free emission. The powerlaw index is also steeper than 2.5, which is expected from an optically thick synchrotron spectrum. 
Hence we conclude that Razin Tsytovich suppression is responsible for this steep spectra.
Earlier works have also come to the same conclusion \citep{bastian01,  carley17}

While good fits were obtained for the vast majority of the spectra, some anomalies deserve mention.
In the 145 MHz map at 03:04 UT, a deep negative was observed at the location of Region 12.
This data point was hence not used for modeling. 
We regard the best fit parameters obtained for this particular spectrum to be unsatisfactory, this is discussed in detail in Sec. \ref{sec:region12}.

\begin{table*}
	\centering 
	\begin{tabular}{p{1.2cm}p{1.8cm}p{1.4cm}c c p{1.8cm}p{1.8cm}c c}
	\hline
	    Region number & Heliocentric Distance & time & E$_{min}^*$& $\delta$ & B & emission area &depth along LOS$^*$&$\chi^2_{reduced}$ \\ 
	    & ($R_\odot$) & (UT) & (keV) & & (G) & (Mm$^2$) & (Mm) & \\ \hline
		7 &2.6&03:32 & 9 & $5.7\pm 0.9$ & $8.7\pm 0.8$ & $48\pm 36$&400 & 1.5 \\ 
		8 &2.7 & 03:32 & 3 & $4.3\pm 0.5$ & $10.5\pm 0.9$ & $43\pm 42$ &400& 1.4  \\ 
		12 & 2.6 & 03:04 & 0.2& $3.6\pm 0.1$ & $15\pm 0.5$ & $705 \pm 300$ &300 & 1.4 \\
		13 & 2.2&03:04 & 3 & $4.4\pm 0.4$ &$12.6\pm 0.4$ &$193\pm 44$&300&0.34 \\
		13 &2.2& 03:18 & 10 & $6.3\pm 0.5$ & $10.8\pm 0.4$& $235\pm 56$&300 &2.2 \\ 
		13 &2.2 &03:33 & 3 & $3.2\pm 0.2$ & $7\pm 1$&$13\pm 7$&300 &1.3 \\ 
		14 & 2.4 &03:04 & 9 & $4.8\pm0.5$ & $9.4\pm 0.5$ &$37\pm 13$ &300& 2 \\ 
		14 &2.4 &03:18 & 3 & $3.14\pm 0.03$ & $7.1\pm 0.3$ &$6^*$&300&4.2 \\ 
		14 &2.4 &03:33 & 9 & $5.7\pm 0.9$ & $9.4\pm 0.9$ &$103\pm 61$&300&0.14 \\ 
		15 &2.5 & 03:04 & 3 & $4.2\pm 0.2$ & $11.4\pm 0.5$ &$50\pm 19$&300&1.5  \\ 
		15 &2.5 & 03:18 & 3 & $4.2\pm 0.4$ & $11.1\pm 0.9$&$67\pm 58$ &300&1.3 \\ \hline

	\end{tabular}
	\caption{Fitted plasma parameters. B denotes the magnetic field. The entries with a $^*$ superscript were kept fixed during the fitting procedure.
	}
	\label{tab:model_spectra}
\end{table*}

\section{Discussion}
\label{Sec:Dicsussion}

\subsection{Radio emission from west limb} \label{sec:radio_CME_discussion}

The only white light feature close to the northern protrusion is the CME.
Given the similarity in morphology and location, it is natural to associate this radio emission with the CME structure.
In the sky plane, the southern protrusion of radio emission of Fig. \ref{fig:avg_normalised} is close to the observed location of a white light streamer.
We examine the possibilities of this radio emission arising from the streamer, interaction between the streamer and the CME, and the CME structure itself.
\begin{figure}
    \centering
    \includegraphics[trim={3.5cm 2.5cm 2cm 2cm},clip,scale=0.5]{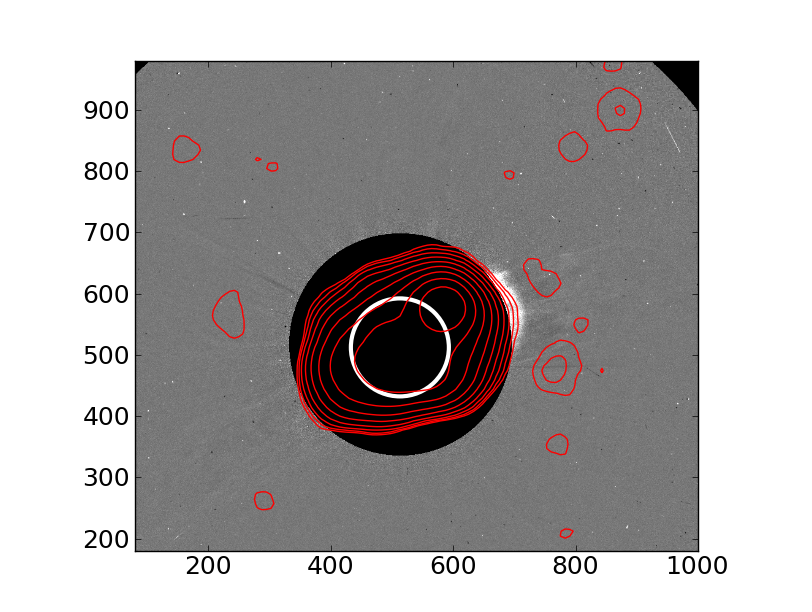}
    \caption{An overlay of 108 MHz radio contours at 02:03 UT on LASCO C2 base difference image at 02:12 UT. 
    The lowest contour is at 0.002 times the peak value in the image, subsequent contours increase in steps of 2. 
    }
    \label{fig:searching_streamer_emission}
\end{figure}

{Figure \ref{fig:searching_streamer_emission} shows an overlay of radio contours on LASCO C2 base difference image at a time the CME enters the C2 field of view. 
While the radio emission from the CME is clearly detected, the emission from the region where the streamer is located is below the detection threshold.
The radio emission appears at the location of the streamer only as the CME passes through this region.

Additional evidence that the radio emission in the southern protrusion cannot be produced by the streamer alone, without any interaction with the CME, comes from the spectrum observed from these regions.}
The relevant mechanism for emission from the streamer is the free-free emission mechanism.
In the Rayleigh-Jeans regime, for an optically thick medium, the emission spectrum goes as $\nu^2$.
As the medium becomes optically thinner, as discussed in Sec. 4.2.2, the spectrum becomes flatter, and under no circumstance, does the spectrum develop any maxima.
All of the modeled spectra show a clear peak, implying that the spectrum is not dominated by free-free emission.

{ Gyrosynchrotron emission, used to model the spectra, requires mildly-relativistic electrons. 
It seems implausible that long lived stable structures, such as streamers, have a steady supply of these energetic electrons.
The absence of radio emission from the streamer prior to the arrival of the CME, and its appearance when the CME is passing through the region, spectra inconsistent with free-free emission, but consistent with gyrosynchrotron emission, and the implausibility of steady availability of mildly relativistic electrons in the streamer, all suggest that the observed emission arises via the  gyrosynchrotron mechanism from the CME plasma.}
 
\begin{figure}
\centering
\includegraphics[scale=0.4]{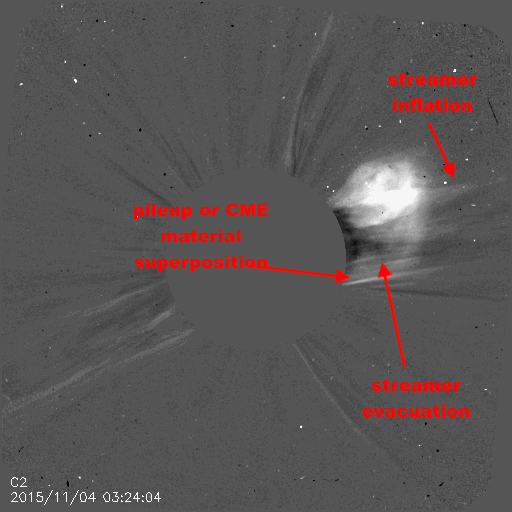}
\caption{Shows a LASCO C2 base difference image showing the observational indication of a CME-streamer interaction.}
\label{fig:CME_streamer_interaction}
\end{figure}

 To examine the possibility of interaction between the streamer and the CME, we closely examine the LASCO C2 difference images.
Fig. \ref{fig:CME_streamer_interaction} shows a typical LASCO C2 base difference image during the course of the CME eruption. The various relevant features are marked on the figure. We clearly see evidence of CME-streamer interaction. Density increase is seen at the edge of the streamer from about 1:25 UT. 
At the northern streamer boundary, this can be clearly associated with the CME-streamer interaction based on the evidence for streamer inflation in the northern side (shown in Fig. \ref{fig:CME_streamer_interaction}). 
However, the situation is uncertain for the southern boundary of the streamer. 
The density enhancement at the boundary can be caused both due to the CME-streamer interaction; mere superposition of the CME and the streamer material in the sky plane due to projection effects; or a bit of both. 
The present data are insufficient to distinguish between these scenarios.

\subsection{Establishing the emission mechanism}
From Fig. \ref{fig:avg_normalised}, it is clear that we can detect the direct radio emission from the CME. Radio emission from the CME has often been associated with plasma emission and free-free emission mechanisms \citep[e.g.][etc.]{gopalswamy1992, ramesh2003}. 
In this section we explore the feasibility of various emission mechanisms, including plasma, free-free and gyrosynchrotron mechanisms, to give rise to the observed emission.

\subsubsection{Plasma emission mechanism}
\label{subsubsec:plasma}
Plasma emission is a coherent emission mechanism and hence results in very high brightness temperatures ($T_B$). 
Even the very weak instances of such emissions have $T_B \gtrsim 10^8 K$ \citep[e.g.][]{Mohan19}, more typical values are at least an order of magnitude higher.
The flux densities observed at the CME flanks correspond to $T_B \sim 10^4 K$, about 4 orders of magnitude lower.
Another key characteristic of plasma emission arising from a homogeneous system is that it is intrinsically narrowband, confined to the local plasma frequency and/or its harmonic.
The vast majority of the emission from the CME flanks lies at heliocentric distances of $\gtrsim 2.8 R_\odot$ and at $\sim 319^ \circ$ ccw. We have estimated the electron density along $319^\circ$ ccw at 2.7 $R_\odot$ to be $1.6\times 10^5$ $cm^{-3}$. This implies that the typical plasma frequencies in these regions is $\lesssim 3$ MHz.
This is much lower than even the harmonic corresponding to our lowest frequency of observation.
At large coronal heights, where this emission is seen, this problem becomes only more acute.


The polarized brightness map used to determine coronal density comes from half an hour earlier.
It is possible that local plasma density has increased at the time of the radio observation due to local instabilities.
To be able to produce plasma emission at our frequencies of observation, the local plasma density needs to increase at least by about a factor of $\sim10^3$ (for emission at the fundamental) and by a factor of $\sim200$ (for harmonic emission) over a large region extending about $0.5 R_{\odot}$ along the LOS and at a heliospheric extent of about $2.5 R_\odot$ in the sky plane. 
It is very unlikely that local random instabilities give rise to an enhancement of such large spatial extent in the sky plane.
The density enhancement cannot be attributed to shock compression as well because the typical density compression ratios attributed to shocks lie in the range of 1--2  \citep{susino15}.

Lastly, if such a density enhancement was to exist, it should be visible as a stark enhancement in intensity in the LASCO C2 images, which is not seen.
Given that the cadence of LASCO C2 images is $12\ min$, and the exposure time for each of the frames is $25\ s$, it is possible to envisage a scenario where these density enhancements took place only during the gaps between successive LASCO observations. 
So while they could not be seen by LASCO, they were visible in the radio observations.
This, however, seems like a contrived and unlikely scenario.
All of these reasons together suggest that the observed radio emission cannot originate from plasma emission mechanism.

\subsubsection{Free-free emission mechanism}
Another possibility worth consideration is that the observed radiation can arise from free-free emission.
Following \citet{gopalswamy1992} and assuming optically thin emission from the CME, the density of CME $n_{CME}$ is related to be brightness temperature of CME by 
\begin{equation}
n_{CME}=( 5\ T_{B,CME}\ T_e^{1/2}\ f^2/L)^{1/2}, \label{eq:density_CME}
\end{equation}
$T_{B,CME}$, $T_e$, $L$ are the CME brightness temperature, electron temperature and LOS depth respectively, and $f$ is the observation frequency.
We find that the estimated number densities using Eq. \ref{eq:density_CME} are much higher than those estimated from the pB map.
For example, the flux density at 108 MHz in Region 2 is $\sim$20 Jy, which corresponds to a brightness temperature of 9609 K.
Assuming a typical line of sight depth of $\sim 0.5 R_\odot$ \citep[e.g.][and this work (Table \ref{tab:model_spectra})]{Tun13,bain14,carley17}, we derive the thermal electron density at Region 2 to be $4\times 10^6$ $cm^{-3}$, which is about 20 times larger than the density estimated in Sec. \ref{subsubsec:plasma}.
Using a smaller LOS depth, or if the total emitting volume is smaller than the size of the PSF, will lead to an even larger density requirement for free-free emission.

We can add another layer of sophistication by taking into account the fact that the pB map was made at 02:58 UT.
Using the speed of the CME (442.3 km/s from CDAW catalogue) and assuming radial propagation, we estimate that the CME would have traveled $\sim 8.99\times 10^5$ km in 33 minutes.
Hence the material residing in Region 2, which lies a heliocentric distance of about $\sim 4.2 R_\odot$, at 03:33 UT would have been at a heliocentric distance of $\sim 2.9 R_\odot$ when the pB observation was made.
The density estimates at these heights also lie in the range of $10^5$ cm$^{-3}$ (Sec. \ref{subsubsec:plasma}).
It should be noted that the density estimated in this way is an upper limit, as the CME also expands in a largely self-similar manner as it propagates out, leading to a decrease in its density.
The arguments in the Sec \ref{subsubsec:plasma} for why instabilities are unlikely to lead to such a density increase also hold true here.

\begin{figure}
    \centering
    \includegraphics[scale=0.35]{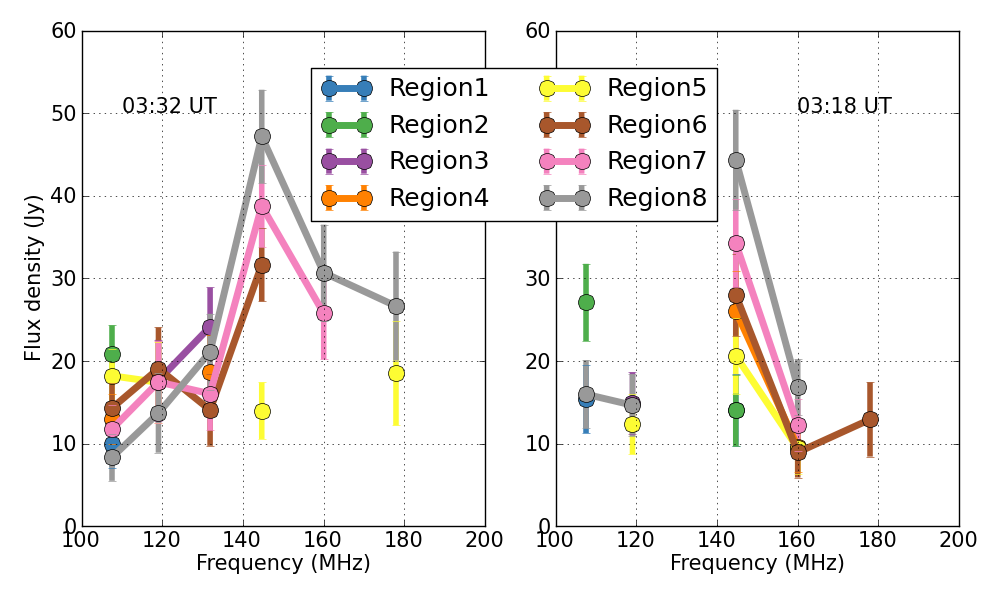}
    \caption{Spectra of regions on northern flank of the CME at 03:18 UT and 03:32 UT are shown.}
    \label{fig:northern_flank_0318}
\end{figure}

{ An examination of the obtained spectra provides stronger evidence against the emission arising from the free-free mechanism.}
As $n_{CME}$, $T_e$ and $L$ are independent of $f$, Eq. \ref{eq:density_CME} implies that $T_B \times f^2 \propto S$ must be a constant.
As $T_B \times f^2 \propto S$, where $S$ is the flux density, this implies that the  flux density must be a constant, independent of frequency.
Figure \ref{fig:northern_flank_0318} shows the spectra of regions on the northern flank at 03:18 UT and 03:32 UT. 
In practically all instances where reliable measurements could be made, the spectra are not flat.
Hence, we conclude that the observed emission is inconsistent with the free-free emission mechanism.


\subsubsection{Gyrosynchrotron emission mechanism}
Having ruled out plasma and free-free emission mechanisms, gyrosynchrotron remains the only likely mechanism for the emission from the flank regions.
In the spectra where sufficient measurements are available, a clear peak, a characteristic of gyrosynchrotron spectra, can usually be identified.
For the cases where flux density estimate could be made only at one or two frequencies, we suspect that the frequencies at which these measurements were made lie close to the peak of the gyrosynchrotron spectrum, and the flux densities fall below our detection threshold at neighbouring frequencies.
These data are, however, insufficient to allow us to confirm this hypothesis by attempting spectral fitting.
\subsection{The curious case of Region 12 at 03:04 UT} \label{sec:region12}
We notice from Table \ref{tab:model_spectra} that the best fit value of $E_{min}$ for Region 12 at 03:04 UT is 0.2 keV, more than an order of magnitude lower than the next smallest value.
It is comparable to the mean electron energy ($\sim 100$eV) for a $10^6$ K corona.
Hence, the assumption made during modeling that the gyrosynchrotron radiation is originating only due to the part of the electron distribution populating the power law, and the thermal electrons do not contribute to it, is no longer valid.
In this instance, the high energy tail of the thermal electrons present in the medium also contributes to the emission, and \citet{fleishman10} implementation of gyrosynchrotron modeling allows this possibility. 

Briefly, in this model, the electron distribution is given by 
\begin{equation}
    n(E)=
    \begin{cases}
    n_{th}(E),& \text{if}\;  E\le E_{cr},\\
    AE^{-\delta},& E\geq E_{cr} \text{ and } E\le E_{max},\\
    \end{cases}
\end{equation}
where $A$ is the normalization constant to ensure that $n(E)$ is continuous at $E_{cr}$.
$E_{cr}$ is defined to be the energy at which $p_{cr}=p_{th}/\epsilon$, where $\epsilon$ is an unknown parameter to be fitted, and $p_{cr}$ and $p_{th}$ are the momenta corresponding to $E_{cr}$ and the mean thermal energy of the electrons.
$n_{th}$ is the number of thermal electrons with energy E.
For small $\epsilon$, $n_{nth}<<n_{th}$.
Hence the normalization of the electron distribution is fixed by $n_{th}$, and $n_{nth}$ is no longer a model parameter.
$E_{min}$ is also no longer a parameter of this model which takes into account the contributions of all electrons present in the system to gyrosynchrotron emission.

\begin{figure}
    \centering
    \includegraphics[scale=0.3]{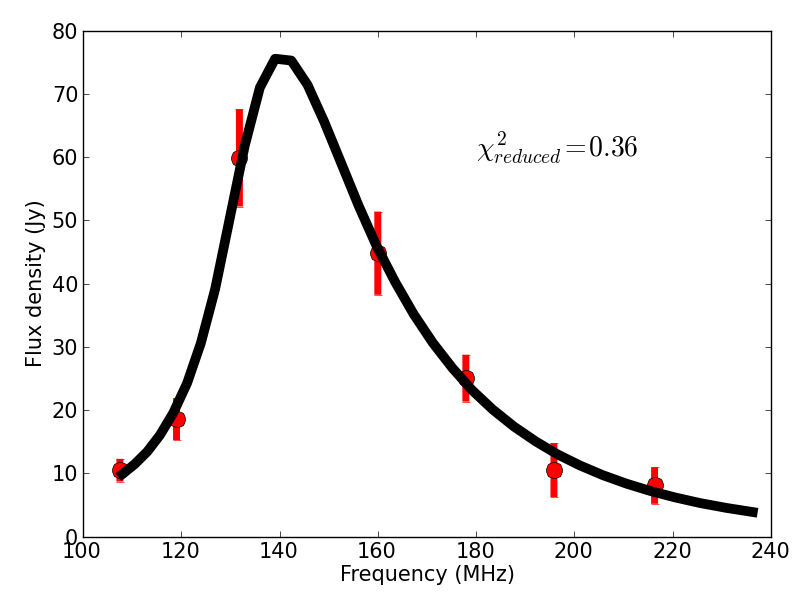}
    \caption{Fitted spectra of Region 12 with both thermal and non-thermal electrons contributing to the gyrosynchrotron emission. The red filled circles denote the data and the black curve is the fitted model. }
    \label{fig:region12_thermal_model}
\end{figure}

The best fit model yielded $\epsilon=0.103\pm 0.002$, $\delta=3.46 \pm 0.09$ and $B=13.7 \pm 0.8$ G, with a $\chi^2_{reduced}=0.36$.
$E_{max}$ was fixed to 100 keV.
Area of emission, line of sight depth and $n_{th}$ were kept at the same values as used in Table \ref{tab:model_spectra}.
Temperature of the region was assumed to be $10^6$ K.
The modeled spectra is shown in Fig. \ref{fig:region12_thermal_model}.
Though some model parameters like the magnetic field and $\delta$ have not changed much, this self-consistent model is much better suited for this scenario.


\subsection{Nonthermal electron energy content} \label{sec:nth}

Here we focus only on the three regions for which we have successfully modeled the spectra at more than one times - Regions 13 through 15.
The energy of nonthermal electrons, $E_{nth}$, is given by
\begin{equation}
    E_{nth}=A\ L\ n_{nth}\ \frac{\delta-1}{\delta-2}\left[\frac{E_{min}^{-\delta+2}-E_{max}^{-\delta+2}}{E_{min}^{-\delta+1}-E_{max}^{-\delta+1}}\right].
    \label{eq:totalE}
\end{equation}
$E_{nth}$ so computed for the three regions is given in Table \ref{tab:energy_content_varying_area}.
\begin{table}
    \centering
    \begin{tabular}{c c c c}
    \hline
    \hline
    & \multicolumn{3}{c}{$E_{nth}\ (\times 10^{14}\ J)$}\\
    \hline
         & 3:04 UT & 3:18 UT & 3:33 UT \\
         \hline
       Region 13  & $11\pm 3$ & $41\pm 8$ & $1.0\pm0.2$ \\ 
       Region 14 & $6\pm 2$ & - & $17 \pm 7$ \\ 
       Region 15 & $3.1\pm 0.4$ & $4\pm 1$ & - \\ \hline
    \end{tabular}
    \caption{Energy content in non-thermal electrons at different regions at different times. Spectra which we have failed to model properly has been left blank. }
    \label{tab:energy_content_varying_area}
\end{table}
In Table \ref{tab:energy_content_varying_area}, we only use spectra for which $\chi^2_{reduced} \le 2.5$. 
The spectrum of Region 14 at 03:18 UT is hence not used ($\chi^2_{reduced}=4.2$). 
For Region 13, $E_{nth}$ first increases by a factor of about 4 over 12 min, and then drops to a small fraction of its peak value in another 15 min.
$E_{nth}$ is essentially unchanged across the two measurements for Region 15, while in view of the large uncertainties for Region 14, the change is not very significant.

Energetic electrons can lose energy in two ways, by collisional and gyrosynchrotron losses. 
Following \citet{tandberg09} and \citet{carley17}, we estimate the collisional loss timescale to be $\sim$15 hours.
Following \citet{takakura1966}, we estimate the gyrosynchrotron loss timescale to be much larger than this ($\sim 10^2$ hours).
The collisional loss timescale is much longer than the timescale at which we observe changes in $E_{nth}$.
\citet{bain14} and \citet{carley17} have also pointed out that the energy loss timescales estimated based on theoretical considerations are much larger than the energy loss timescale estimated from the data.
The observed increase in the total non-thermal energy is also found to be accompanied by an increase in the absolute number of non-thermal electrons.
The observed variability in $E_{nth}$ and bulk electron content implies that not only the nonthermal electrons are able to stream through the CME plasma, but the existence of a mechanism for generation of sufficiently energetic nonthermal electrons $\sim$ 2 hours after the initial eruption.

\begin{figure*}
     \centering
     \gridline{\fig{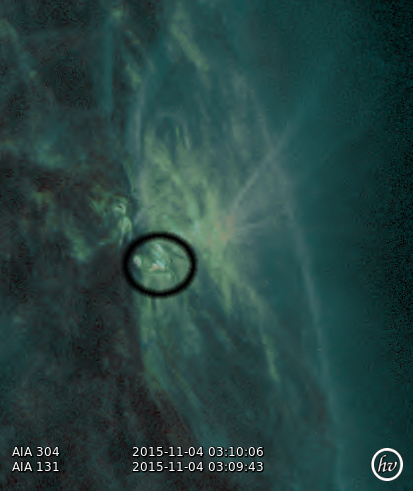}{0.19\textwidth}{03:10}
           \fig{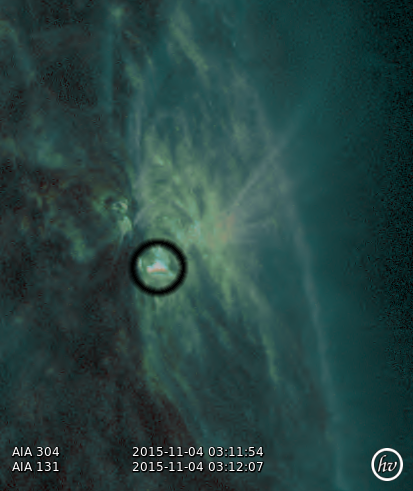}{0.19\textwidth}{03:12}
           \fig{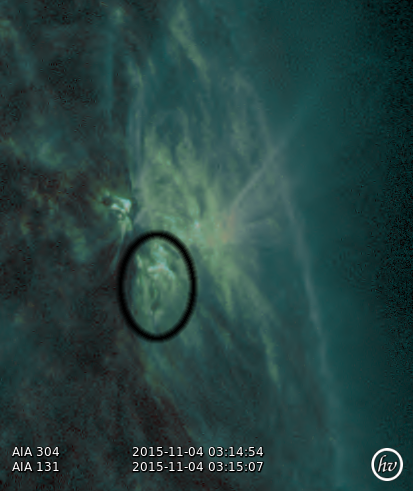}{0.19\textwidth}{03:15}
           \fig{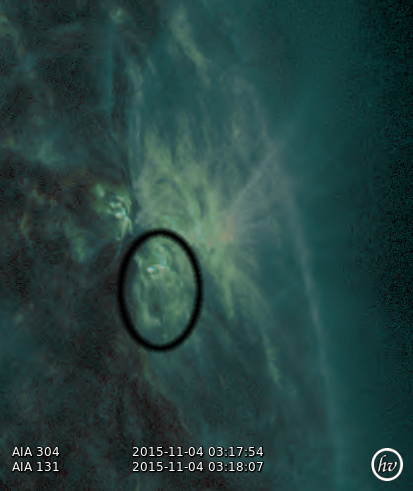}{0.19\textwidth}{03:18}
           \fig{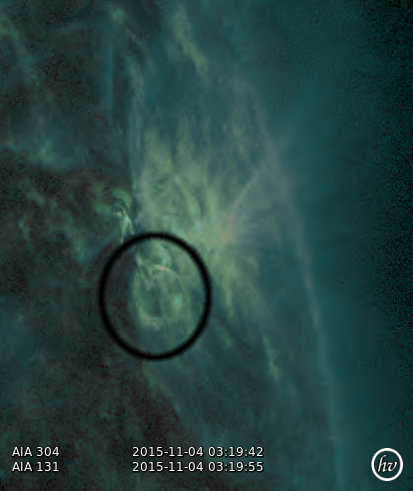}{0.19\textwidth}{03:20}}
     \caption{A series of AIA 131 \AA\ and 304 \AA\ images highlighting the late time brightenings. The black circles denote the region of interest.}
     \label{fig:late_brightenings}
 \end{figure*}

This suggests that the small scale reconnections, responsible for injecting these energetic particles, continue to take place even at late times. 
This hypothesis is also supported by the multiple brightenings seen in AIA 131~\AA, 171~\AA, 304~\AA\ images around 03:12 UT and material outflowing from the active region site at around 03:18 UT.
Fig. \ref{fig:late_brightenings} shows some of these EUV brightenings.

It is reasonable to expect radio signatures of these small reconnection events, e.g. type I noise storm or type III radio bursts.
While no type III bursts are observed  in the MWA bands, a compact nonthermal source is visible co-located with the EUV brightenings, and might correspond to the type I noise storm source.
It is possible that even if type III emission is produced, it lies outside the MWA observing band.
There are also reasons which suggest that the origin of the type IIIs themselves might be suppressed - they range from continuous injection of energetic particles, which seems to be the case here \citep{reid2014} to presense of density inhomogenities in the CME plasma \citep{kontar2009,reid2013,reid2014}.

\subsection{Comparison with previous works}

In their quest to cover large spectral ranges, some of the earlier works combined data from single dishes, which cannot provide any spatially resolved information, to complement the spatially resolved information from imaging instruments like the Nan\c{c}ay Radio Heliograph \citep[NRH,][]{kerdraon1997} \citep[e.g.][]{maia07,carley17}. 
While driven by necessity, this approach had some significant limitations.
In absence of imaging, an average pre-burst flux density was subtracted to arrive at an approximate estimate the nonthermal radio flux from the CME. 
Perhaps more importantly, the spectral modeling required them to assume the source to be homogeneous, even though the importance of spatially resolved observations and the need for inhomogeneous gyrosynchrotron models was already recognised \citep{klein1984}.


Some other studies relied exclusively on spatially resolved spectra obtained from the NRH \citep{bastian01, Tun13, bain14}. 
Due to the combined effect of the spectral coverage of the NRH (150--450 MHz) and the nature of the spectra observed, these spectra rarely sampled the peak and the low frequency part of the gyrosynchrotron spectrum.
This diminished the ability of these measurements to constraint the fit parameter.

\begin{figure}
    \centering
    \includegraphics[scale=0.4]{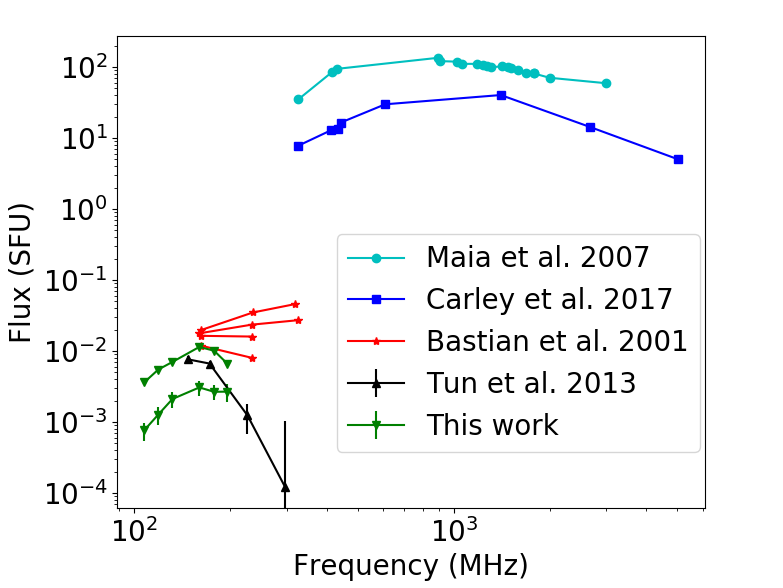}
    \caption{Comparison with previous works}
    \label{fig:past_work_comparsion}
\end{figure}

The MWA operates at a comparatively lower part of the band, best suited for observations of gyrosynchtoron emission from CME plasma at larger heliocentric distances, and longer durations after the launch of the CME.
The much higher imaging dynamic range provided by the MWA enables us to reliably estimate the lower flux densities to which these emissions fall by the time the peak of the spectrum moves into the MWA range.
Additionally, the denser MWA spectral sampling helps with being able to constrain the multiple free parameters of a gyrosynchrotron model.
Figure \ref{fig:past_work_comparsion} provides a compilation of all of the past works using spatially resolved spectra of gyrosynchrotron emission associated with CME plasma\footnote{A spectrum is regarded as spatially resolved only if all data points on it came from an image.}.
Two example spectra from this work are also included to illustrate the points mentioned above.
Table \ref{tab:past_work_comparison} compares some of the gyrosynchrotron model parameters derived from this work with those from earlier works. 

We note that while the parameters from \citet{bastian01} and \citet{maia07} come from a structure which closely resembles the white light CME, other works have concentrated on the CME core \citep{Tun13,bain14,carley17}. 
In this work, we have modelled some regions from the southern flank of the CME, and some regions lying between the CME flanks that maybe the counterparts of a core, if such structure was detected in the white light images.

\begin{table*}[ht]
    \centering
    \begin{tabular}{c p{1.3cm} p{1.3cm} p{1.0cm} c c p{1.3cm} p{2.0cm} c}
    \hline
         	Reference & \begin{tabular}{c}E$_{min}$\\(MeV)\end{tabular} & \begin{tabular}{c}E$_{max}$\\(MeV)\end{tabular} & $\delta$ & B(G) & $n_{nth}$ & \begin{tabular}{c}Distance\\($R_\odot$)\end{tabular}& Time elapsed after flare(min) & Resolved ?\\\hline\hline
         	\citet{bastian01}& 0.1 & 10 & 3.5 & 0.33-1.47 & 2$\times 10^2$ & 1.45-2.8 & 56 & Yes\\ 
         	\citet{maia07} & 1-6 & - & 1.5,3.5; Not much dependence found. & 0.3-8 & - & - &7-8 & No \\ 
         	\citet{Tun13} & 0.001 & 0.1 & 3 & 6-23 & 3$\times 10^5$-2$\times 10^6$ & 2-2.5 & 37 & Yes \\ 
         	\citet{bain14} & 0.01-0.316 & 10 & 5 & 3.7  & 3.98$\times 10^4$ & 2-2.5 & 37 & Yes \\ 
         	\citet{carley17} & 0.009 & 6.6 & 3.2 & 4.4 & $10^6$ & 1.3 & 4 & No \\ 
         	Region 8 (This work) & 0.003 & 10 & 4.3 & 10.5 & 3$\times 10^4$ & 2.73 &121 & Yes \\ 
         	Region 13 (This work) & 0.003 & 10 & 3.2 & 7 & 3$\times 10^4$ & 2.3 &122 & Yes \\ \hline
    \end{tabular}
    \caption{Comparison of fitted parameters with previous works. }
    \label{tab:past_work_comparison}
\end{table*}

Though not mentioned in the table, the LOS depth used in this work is similar to previous works.
One difference between this work and the previous ones is the area of emission.
The area of emission is not provided by the previous works modeling the radio CMEs \citep{bastian01,maia07}.
Due to lack of a better option, we compare the emission area estimated here to that from other works where the observed radio emission does not follow the CME morphology \citep{Tun13,carley17,bain14}.

Where the earlier authors have provided the emission area, it is $\sim10^{20}$ cm$^2$, which also happens to be close to the size of the PSF of instruments used, and that of the MWA for that matter.
We use a much smaller emission area, in the range $\sim10^{17}$--$10^{18}\ cm^2$.
Given this large difference between all earlier works and this one, we investigated if there exists a region in the parameter space where an emission area in the vicinity of the conventional choice provides a good fit to the observed spectrum.
For this, we choose to construct model spectra across a large span of parameter space and compute its $\chi^2$ with respect to the observed spectra.
The various parameters which were varied, with their ranges of variation and the step sizes mentioned in parenthesis, were: emission area ($1$--$10 \times 10^{20}$ cm$^2$ in steps of $3 \times 10^{20}$ cm$^2$), 
$E_{min}$ ($0.1$--$7$ keV in steps of 0.5 keV),
$E_{max}$ ($0.5$--$10$ MeV in steps of 2.0 MeV),
$\delta$ ($0.5$--$9$ in steps of 0.5),
magnetic field ($0.4$--$5.0$ G in steps of 0.3 G), and
$n_{nth}$ ($10$--$5000$ cm$^{-3}$ in steps of $100$ cm$^{-3}$ ).
The minimum value of $\chi^{2}$ ($\chi^{2,min}$) was found to be 170.4, whereas the largest $\chi^{2,min}$ for the fits shown in this paper range is $\sim 20$.
The corresponding parameter values were emission area = $10^{20}$ cm$^2$, $E_{min}=0.1$ keV, $E_{max}=8.5$ MeV, $\delta=2.0$, $B=0.4$ G, $n_{nth}=710$ cm$^{-3}$.
It is immediately obvious that the $\chi^{2,min}$ is much larger than that obtained by assuming much smaller values for emission area.
Additionally, $\chi^{2,min}$ is obtained not only at the lower limit of emission area, but also the lower limit for $E_{min}$ and $B$ in the exploration grid.
This suggests that the true $\chi^2$ minima lies below 0.1 keV.
However, 0.1 keV is already the average energy of thermal electrons in a $10^6$ K corona and a lower $E_{min}$ would be aphysical.
Hence, the observed spectra are not consistent with a large emission area $\sim10^{20}\ cm^{2}$.

\subsubsection{Interpreting the small emission area}

We consistently find that the best fit emission area is about a few percent of the PSF area (Table \ref{tab:model_spectra}).
So while the emission is seen to be filling the entire synthesised beam, the modeling, which has no information about the PSF, insists that it must come from within a tiny fraction of the PSF.
There are only two possible ways which allow the emission area to be this small. 
Either the nonthermal electrons have a very small effective filling factor, or the emission is arising from regions where the magnetic field is concentrated into small knots with very strong magnetic fields, or perhaps a mix of both. 
Magnetohydrodynamic simulations of CMEs shows the existence of magnetic knots through out the magnetic flux rope \citep{karpen2012}.
The possibility of a small filling factor of nonthermal electrons can be explained in the following manner.

For impulsive injections, the large range of energies spanned by the electron distribution power law, leads to a corresponding spread in positions of electrons of different energies as they travel along magnetic field lines.
For instance, a bunch of electrons with energies ranging between 1--100 keV released instantaneously at the solar surface spreads itself over a linear dimension 1.2 $R_{\odot}$ by the time 50 keV electrons have traversed $1\ R_{\odot}$.
Though collectively these electrons occupy large regions, the velocity dispersion implies that at any given time, a small part of this region is populated by electrons from only a small range of the initial energy distribution.
This is not taken into account by the gyrosynchrotron modeling framework, which expects each region to be populated by the electrons representing the entire distribution.
This dramatically reduces the effective emission area, as estimated by the gyrosychrotron spectral modeling framework.

\section{Conclusion}
\label{Sec:Conclusion}

We have presented spatially resolved observations of gyrosynchrotron emission from CME plasma.
The DR of images used in this work typically is 13,000, much higher than previous works. 
The availability of these high DR images is the primary reason which has enabled this detailed investigation into the problem of CME radio emission.
Multiple aspects of this work represent a significant advance over earlier studies.
These include - the lowest reliable flux densities reported; and both the farthest heliocentric distances and late times at which CME radio emission has been detected yet.
The fine spectral sampling of these data allow us to carry out a robust determination of the CME magnetic fields and other interesting parameters under some plausible assumptions.
We report an instance where the usual assumption that thermal electrons do not play a role in gyrosynchrotron emission is violated.
The energy content of nonthermal electrons in different regions is found to vary over time scales of minutes, even at late times after the CME eruption, suggesting sustained but sporadic particle acceleration processes either at the CME site or at the shock front giving rise to these nonthermal electrons.

Our estimates of the emission area of gyrosynchrotron emission ($\sim 10^{18} cm^2$) are much lower than those reported in past works ($\sim 10^{20} cm^2$). 
We interpret this in terms of a very low filling factor of the nonthermal electrons, leading to an effectively much smaller emission area.

The focus of this study was a relatively slow CME with a speed of $442.3\ km\ s^{-1}$.
It may have driven a weak shock and was likely accompanied by comparatively less energetic populations of nonthermal electrons, as compared to faster CMEs.
A radio noise storm was also in progress close to the site of eruption.
Despite these unfavourable conditions, we have successfully detected gyrosynchrotron emission from a CME at six frequencies spanning 108--220 MHz out to a height of 2.7 $R_\odot$, at lower flux densities than have been reported before.
This suggests the imaging quality achieved here should be quite sufficient to routinely detect radio emission from CME plasma using data from the MWA.
Work is currently underway to obtain well calibrated Stokes V maps, which will further improve our ability to model gyrosynchrotron spectra.
It is important to bear in mind, however, that a detailed modeling of the spectrum requires additional information beyond what can be obtained from radio observations alone.
In particular, the radio observations could be significantly augmented by off-limb EUV and FUV spectroscopic observations \citep[e.g.][]{laming2019}. The combination would permit determination of the magnetic field and energetic particle content of CMEs early in their evolution with likely important implications for  Space Weather prediction ability.

\acknowledgments
This scientific work makes use of the Murchison Radio-astronomy Observatory (MRO), operated by the Commonwealth Scientific and Industrial Research Organisation (CSIRO).
We acknowledge the Wajarri Yamatji people as the traditional owners of the Observatory site. 
Support for the operation of the MWA is provided by the Australian Government's National Collaborative Research Infrastructure Strategy (NCRIS), under a contract to Curtin University administered by Astronomy Australia Limited. We acknowledge the Pawsey Supercomputing Centre, which is supported by the Western Australian and Australian Governments. 
The SDO is a National Aeronautics and Space Administration (NASA) spacecraft, and we acknowledge the AIA  science team for providing open access to data and software. 
The SOHO/LASCO data used here are produced by a consortium of the Naval Research Laboratory (USA), Max-Planck-Institut f\"ur Aeronomie (Germany), Laboratoire d$^{’}$Astronomie (France), and the University of Birmingham (UK). SOHO is a project of international cooperation between the ESA and NASA. 
This research has also made use of NASA's Astrophysics Data System (ADS). 
We also thank the staff of Learmonth spectrograph for making their data public.
SM also gratefully thanks Karl-Ludwig Klein (LESIA-Observatoire de Paris) for his detailed comments on an early draft of this work, which led to significant improvements. We also thank Apurba Bera (NCRA) and Devojyoti Kansabanik (NCRA) for useful discussions. 
AV is supported by NRL grant N00173-16-1-G029 and NNX16AG86G.
SM and DO acknowledge support of the Department of Atomic Energy, Government of India, under the project no. 12-R\&D-TFR-5.02-0700.

\facility{Murchison Widefield Array, SDO (AIA), RHESSI, GOES, SOHO, Learmonth Spectrograph}

\software{CASA,
SolarSoft Ware,
Python 2.7\footnote{\href{https://docs.python.org/2/index.html}{https://docs.python.org/2/index.html}},
NumPy\footnote{\href{https://docs.scipy.org/doc/}{https: //docs.scipy.org/doc/}},
Astropy\footnote{\href{http://docs.astropy.org/en/stable/}{http://docs.astropy.org/en/stable/}},
Matplotlib\footnote{\href{http://matplotlib.org/}{http://matplotlib.org/}}}

\bibliography{bibliography}



\end{document}